# Design and testing of an sTGC ASIC interface board for the ATLAS New Small Wheel upgrade

Xu Wang[a,b,c] Liang Guan[c], Siyuan Sun[c], Bing Zhou[c], Junjie Zhu[c] and Ge Jin[a,b]

[a]State Key Laboratory of Particle Detection and Electronics, University of Science and Technology of China, Hefei, 230026, China
[b]Department of Modern Physics, University of Science and Technology of China, Hefei, 230026, China
[c]Department of Physics, University of Michigan, Ann Arbor, MI, 48109, USA

ABSTRACT

The ATLAS experiment will replace the present Small Wheel (SW) detector with a New Small Wheel detector (NSW) aiming to improve the performance of muon triggering and precision tracking in the endcap region at the High-Luminosity LHC. Small-strip Thin Gap Chamber (sTGC) is one of the two new detector technologies used in this upgrade. A few custom-designed ASICs are needed for the sTGC detector. We designed an sTGC ASIC interface board to test ASIC-to-ASIC communication and validate the functionality of the entire system. A test platform with the final readout system is set up and the whole sTGC readout chain is demonstrated for the first time. Key parameters in the readout chain are discussed and the results are shown.

## 1. Introduction

The ATLAS detector [1] is one of the two general-purpose particle detectors at the Large Hadron Collider (LHC) at CERN [2]. To increase the potentials for new discoveries, the LHC will be upgraded to the High-Luminosity LHC (HL-LHC) with the instantaneous luminosity increased from $10^{34}cm^{-2}s^{-1}$ to $5 - 7.5\times 10^{34}cm^{-2}s^{-1}$ [3]. ATLAS will replace the present small wheel muon detector，the innermost endcap station, with a New Small Wheel (NSW) detector [4] to handle the harsh conditions expected at the HL-LHC.

The NSW detector employs two detector technologies: resistive Micromesh Gaseous Structure detector (MicroMegas) [5] and small-strip Thin Gap Chamber (sTGC) [6], both providing triggering and precision tracking for muons. It will provide online segment measurements with an angular resolution of ~1 mrad and offline hit position measurements with a resolution of ~100 microns. A NSW sector consists of 16 detector planes with a sandwich arrangement of sTGC-MM-MM-sTGC. The sTGC is a multiwire proportional drift chamber. The basic detector structure consists of a grid of gold-plated tungsten anode wires sandwiched between two resistive cathode planes. One cathode is covered with ~8×(8-60) $cm^2$ pads, while the other cathode is covered by strips with 3.2 mm in pitch and 0.5 - 2 m in length. The pads are used to identify regions of interest for muons produced from the interaction point. The precision hit positions are determined by charges collected by the strips.

---

\* Corresponding authors.
  E-mail address: waxu@umich.edu (Xu Wang).

The sTGC trigger and readout system is shown in Figure 1. Raw detector signals are first amplified, shaped, and discriminated by a VMM ASIC [7-9]. This ASIC provides precise amplitude and timing measurements as well as prompt trigger signals for 64 detector channels. A Trigger Data Serializer (TDC) ASIC [10-11] accepts VMM signals and prepares trigger data for both strips and pads, performs pad-strip matching using regions of interest provided by the Pad Trigger board, and serializes the strip charge data to the Router board. The Router board [12] collects TDS output data and sends them to the sTGC trigger processor [13] via optical fibers. The sTGC trigger processor performs segment-finding and determines the direction of track segments. Segments found by the NSW and the Big Wheel (the endcap middle station) are combined to form a muon track inside the Sector Logic [14] for muon triggering.

Two stages of trigger signals, Level-0 (L0) and Level-1 (L1), and additional L0 data buffering inside the VMM are needed in the sTGC readout chain in order to be compatible with the readout scheme of the ATLAS experiment. After receiving the L0 accept (L0A) trigger signal, VMM will transmit the L0 data to a third ASIC called ReadOut Controller (ROC) [15-16]. ROC performs trigger matching after receiving the L1 accept (L1A) trigger signal and transmits the L1 data to an sTGC Level-1 Data Driver Card (L1DDC) [17]. The L1DDC then forwards the data to a network called Front End LInk eXchange (FELIX) [18]. Two ASICs designed by the CERN microelectronics group, GigaBit Transceiver (GBT) [19] and Slow Control Adapter (SCA) [20], are used for data transmission and system configuration. The total number of readout channels for the entire sTGC detector is 354k.

It is particularly important to assemble them on the same board and perform integration studies before we finalize the ASIC designs even if various test fixtures have been designed to perform thorough tests of these custom-designed ASICs. We present in this article the design of an sTGC ASIC interface board with all ASICs mounted and detailed studies we conducted. This board allowed us to perform a system-level test of these ASICs for the first time. Many test points were provided so that we can probe some key signals and obtain a better understanding of the behaviors of these chips. This board also acted as the first mini-prototype board for the final sTGC front-end boards (FEBs) [21].

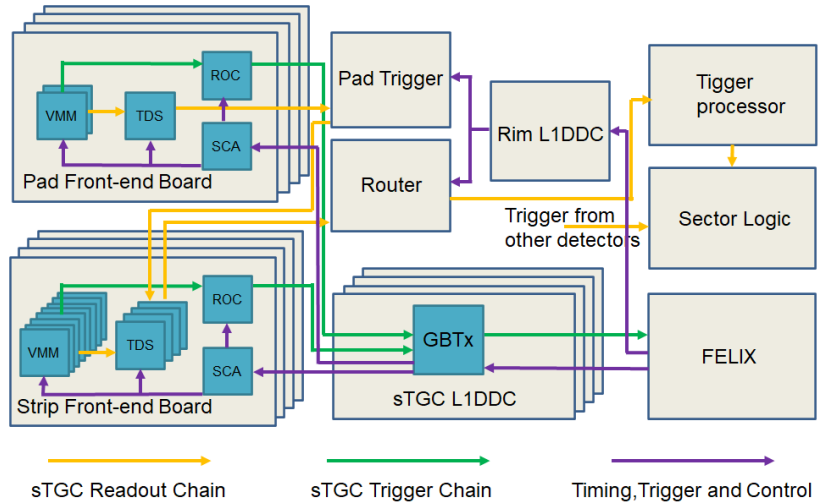

Figure 1. Schematic diagram of the sTGC trigger and readout system.

## 2. Hardware

Figure 2 shows the schematic diagram of the sTGC ASIC interface board. In total five ASICs (two VMMs, one TDS, one ROC, and one SCA) are placed on the board. Two VMMs are used since we expect to have one TDS receive data from two VMMs in the final configuration. VMM processes both pad and strip detector signal and digitizes the signal's peak amplitude and time using a 10-bit ADC and an 8-bit TDC respectively. The L0 data, tagged by Bunch Crossing identifier (BCID), is transferred to the ROC at a rate of 640 Mbps per VMM after the VMM receives the L0A signal. The ROC aggregates the precise amplitude and timing L0 data from up to eight VMMs. It then performs trigger-matching after receiving the L1A signal and serializes the data to the DAQ system. In addition, the VMM provides the time-over-threshold signal for each pad and the fast 6-bit ADC amplitude information for each strip to the TDS at a rate of 320 Mbps per channel for triggering purposes.The TDS receives trigger data from up to 104 pad channels or up to 128 strip channels, performing pad-strip matching, and transmitting trigger data for the selected strips at a rate of 4.8 Gbps.

The SCA is used to configure and distribute clock [22] signals to other ASICs. The SPI (Serial Peripheral Interface) protocol is used to configure the VMM, a 7-bit I2C (Inter-Integrated Circuit) protocol is used to configure the TDS, and a 10-bit I2C protocol is used to configure the ROC. The SCA also controls the mode select and reset other ASICs with GPIO (general-purpose input/output) pings. The connections between ASICs and data-transfer speeds for important data and clock lines are also listed in Figure 2.

Three MiniSAS connectors (MiniSAS 1-3) and one 300-pins GFZ connector are placed on the board to simulate the final application. MiniSAS1 receives 40 MHz bunch crossing (BC) clock and TTC signals and sends them to the ROC. The ROC decodes and forwards TTC information and also provides clock signals to the other frontend ASICs. The ROC output will be sent out through MiniSAS1. In addition, MiniSAS1 connects the SCA to the control system to receive configuration commands and send replies. MiniSAS2 transmits the TDS output at 4.8 Gbps, and MiniSAS3 receives the trigger input for the TDS. The GFZ connector is used to inject external signal pulses to the VMMs.

Figure 3 shows the top and bottom views of the actual interface board. The board is 17 cm long and 6 cm wide, similar to the size of the final front-end board. Wire-bond Ball Grid Array packages are used for all five ASICs on the board. The VMM and the TDS have 400 pins, the ROC has 256 pins and the SCA has 196 pins. The board has fourteen layers total including six signal layers, two power layers and six ground layers to make high-density connections among these ASICs possible. The board also features many test points for debugging and monitoring of important signals.

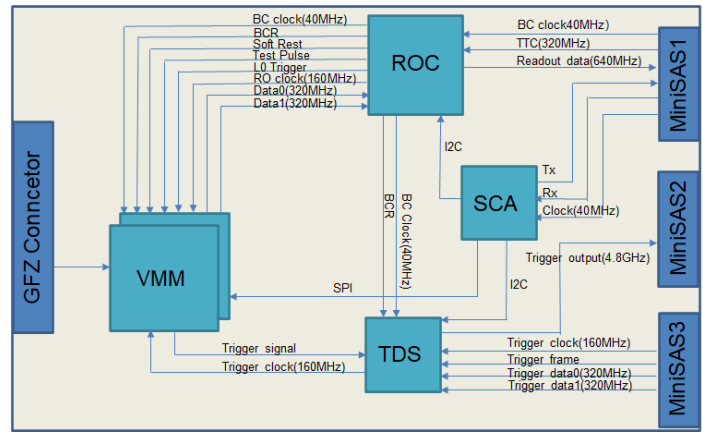

Figure 2. Schematic diagram of sTGC ASICs interface board.

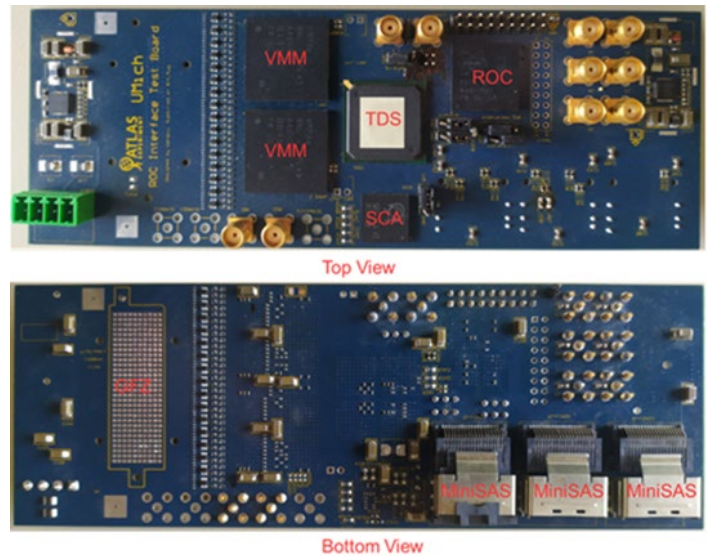

Figure 3. Top and bottom views of the actual sTGC ASIC interface board.

## 3. Test platform setup

### 3.1. Test platform

Figure 4 shows the layout of the test platform we built. FELIX acts as a network server, sending commands to the frontend electronics and broadcasting any received replies and data to the commercial network. Control, configuration, TTC and clock signals are sent to the interface board via the L1DDC. In addition, the L1DDC aggregates and transmits the ROC output data from multiple FEBs to FELIX. MiniSAS cables are used to connect the L1DDC and the sTGC ASIC interface board. A scope is used to monitor the analog output and probe on test points for a better understanding of chip behaviors. Data is passed through the entire sTGC readout chain (VMM → ROC → L1DDC → FELIX) to reach the PC.

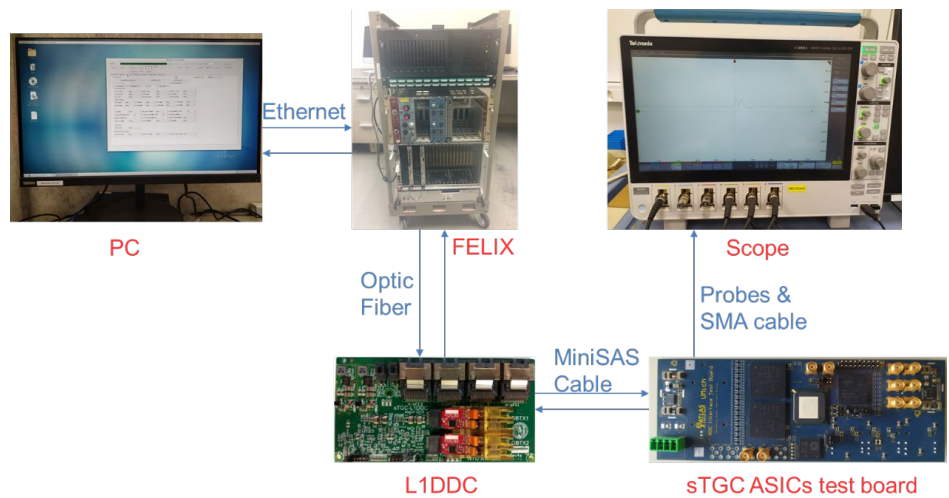

Figure 4. Block diagram of the mini-DAQ system.

### . 3.2. Generating the TTC signals

The TTC signal is a serial signal running at 320 Mbps. It sends out an 8-bit trigger type word in each BC (25 ns): L0A, L1A, test pulse (TP), L0 event counter reset (EC0R), L1 event counter reset (ECR), soft reset, Bunch Counter Reset (BCR), and SCA reset. The ROC samples the TTC signal and distributes the information to other ASICs.

It is critical to tune the phases of the 40 MHz and 160 MHz clocks provided by the ROC in order to have the eight TTC bits decoded correctly. We send one TTC bit at a time and check the presence of the corresponding TTC output to tune the clock phases. If we have access to the VMM analog output, we can also send the TP bit and check the analog output using a scope. In the case that we cannot access the VMM analog output, we can send the L1A bit and check the appearance of the L1 data packet.

The TTC system is programmable and we can define the pattern of the TTC signals and adjust relative latencies between these signals. We first issue a BCR and a OCR (Orbit Counter Reset, which is two consecutive BCRs) to reset the system, and then we send out a TP command so that the VMM internal test pulser will generate a TP signal. The signal will be processed and digitized data is stored in the VMM buffers. After that we send the L0A and L1A signals to read out data from the VMM and the ROC. Figure 5 shows the screenshot of the TTC signal generated for the readout test.

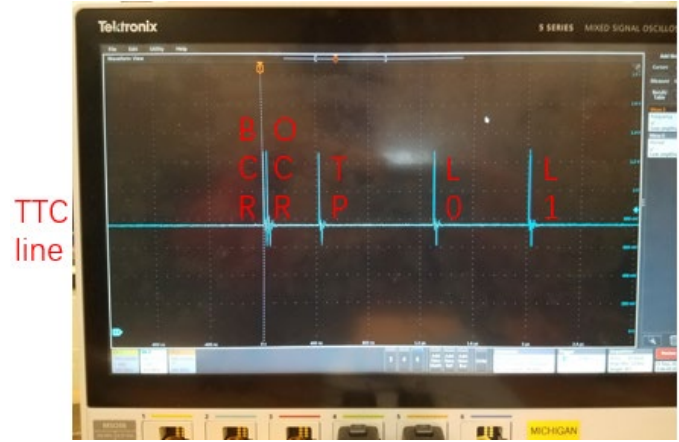

Figure 5. Sequence of the TTC signal shown on the scope.

### .3.3  Test procedure for the sTGC readout chain

There are several readout modes implemented inside the VMM and the so-called Level-0 mode is used by default. This mode allows the VMM to buffer the channel hit data for a period of 16 μs and to select the data to be read out after the reception of the L0A signal at a frequency up to 4 MHz [4]. Data for each hit contains a 12-bit BCID for the timestamp, a 10-bit word for the signal amplitude, and a 8-bit word for the timing. When a L0A signal arrives, the VMM performs trigger matching with a programmable trigger window of up to 8 BCs. Only hit data for which their timestamp is within the trigger window are copied to the VMM L0 channel FIFO and then packaged to form the L0 data by the L0 Event Builder. The L0 data packets are encoded to the 8b/10b format [23] for data transmission between VMM and ROC via two serial data lines (even bits on one line and odd bits on the other). Each data line is driven by the 160 MHz readout clock from ROC and runs with Double Data Rate (DDR), the overall data rate between VMM and ROC is thus 640 Mbps. After receiving a L1A signal, a second-stage trigger matching is performed by the ROC to build the L1 event packet.

To read out the sTGC data, we first set up the FELIX and L1DDC system and make sure we can configure all chips and distribute TTC patterns correctly. Then we train the GBTx ASIC on the L1DDC with the comma pattern sent by ROC and ensure the data transaction from the ROC to the L1DDC is stable. After that, we tune the ROC-to-VMM clock phases to ensure the data transmission from the VMM to the ROC is correct. We first check the L0 data by setting the ROC to the bypass mode so that the L1 trigger matching inside the ROC is bypassed and the L0 data packets are sent directly to FELIX. After we fully understand the L0 data, we configure the ROC to perform the L1 trigger matching and send out the L1 data packets. Finally, we send test pulses and analyze the output data from the entire sTGC readout chain.

## 4. Test platform setup

### 4.1  *Clock phases*

The sTGC readout chain test involves lots of signal transmission between different electronics components and adjustments of relative phases of many clock signals. For the reception of the TTC signal, the ROC internal 40 MHz and 160 MHz clocks need to be aligned to decode the TTC bits correctly. For the L0 data transmitted from the VMM to the ROC, the 160 MHz readout clock phase needs to be set properly to avoid transmission errors. Comma signals are constantly sent on the data lines when there is no trigger signal. The status of the data link can be inferred from ROC registers. If the link status shows errors, the readout clock phase needs to be adjusted. For the ROC output data to the L1DDC, there is also a phase tuning needed for the clock signal on the L1DDC side.

### 4.2  *L0 matching*

After the clock phases are tuned, we start with the L0 matching in the VMM. The BCID of the L0A signal and the BCID of the hit data need to be within the trigger-matching window for the L0 matching. When a pulse arrives at VMM, its BCID is recorded and added to the data packet. When the L0A signal arrives, the L0A BCID is also recorded and the L0-matching algorithm is applied. All hits stored will be checked and only those hits inside the trigger-matching window will be read out. Since VMM has a dead time that is longer than the maximum trigger-matching window, there won't be multiple hits stored in one channel in each data packet.

In order to synchronize electronic systems for different detectors, an offset between the hit BCID and the L0A BCID is introduced to adjust latencies for different electronic systems. The hit BCID can be tuned by the channel-tagging BCID offset, and the L0A BCID can be tuned by the L0 BC offset. The actual BCID for an event is the time difference between BCR and TP plus the channel-tagging BC offset, while the actual L0A BCID is the time difference between BCR and L0A plus the L0 trigger offset. To have successful L0 matching, we can either tune the latency between the TP signal and the L0A signal on the TTC line or tune three BCID-related VMM configuration parameters: the L0 BC offset, the channel-tagging BC offset, and the BC rollover.

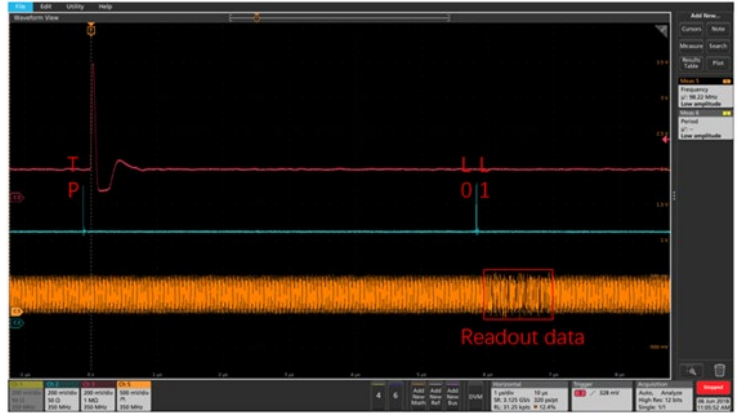

Figure 6. L0 data on the scope.

We use a probe to monitor the data on the VMM data lines. Figure 6 shows an example. The TP signal sent by the TTC and the actual pulse after the VMM amplification can be seen on the analog output. The L0A and L1A signals are sent after the TP, and the L0 data are shown as the orange line in the red box.

### 4.3  L1 matching

Exact matching of the BCID and the orbit ID of the L1A and the L0 data packets are required for the L1-matching in the ROC. The ROC assigns the L1A BCID. The actual L1A BCID is the time difference between BCR and the time at which the L1A arrives plus the L1 BCID offset. The orbit ID for the L1A signal is a 2-bit counter recording the number of rollovers for the BCID. It is needed because the ATLAS L1 trigger latency is more than one rollover. The orbit ID is added to the L0 data packet but not used in the L0-matching algorithm. For simplicity's sake in our test, we sent the L0A and L1A signals in the same BC and set the L0 and L1 BC offers to be the same.

A ROC ASIC takes output from a maximum of eight VMMs. The L0 data are deserialized first and then decoded from the 8b/10b format and stored in FIFOs. A multiplexer is used to assign eight inputs to four separate sub-ROCs (sROCs) according to the input data rate to balance the input data rate of each sROC. Each sROC performs the L1-matching algorithm and builds its L1 data packet. The L1 packet will be encoded to the 8b/10b format and be serialized. The data speed for each sROC can be configured to 80 Mbps, 160 Mbps, 320 Mbps or 640 Mbps. Outputs from all four sROCs are combined to form an event.

### 4.4  Test results

After we validated the L0-matching and L1-matching data from the VMM and the ROC, we varied the amplitude of the test pulse by configuring the pulser DAC in VMM and took data for different DAC values. We developed a C++ program to decode all data packets and obtain ROC ID, L1 event ID, orbit ID, BCID, VMM ID, channel number, relative BCID, 10-bit signal amplitude measurement and 8-bit timing measurement for each hit.

Figure 7 shows the average signal amplitude as a function of the input test pulser DAC value for eight VMM channels. The dependence is found to be linear for DAC values below 500 and above that value, the signal amplitude is saturated around 900 as expected.

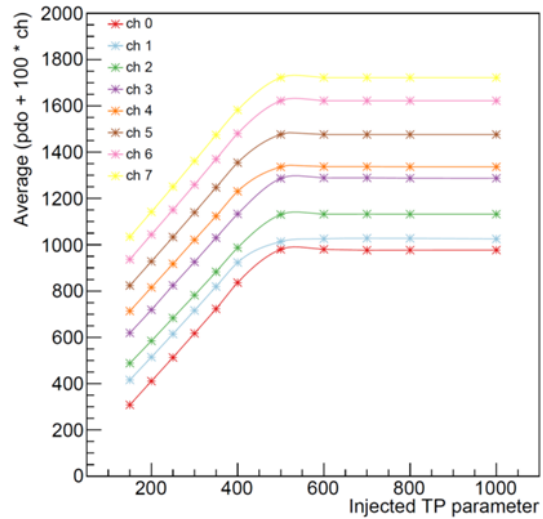

Figure 7. Average signal amplitude as a function of the input test pulser DAC value for eight VMM channels. The average signal amplitude is increased sequentially by 100 according to the channel number so that the eight curves are separated.

## 5.    Summary

We designed an sTGC ASICs interface board and conducted the first study of the readout chain using custom-designed ASICs from the ATLAS collaboration for the NSW upgrade. We demonstrated stable communications between these ASICs. We also demonstrated the entire sTGC readout chain with a L1DDC and a FELIX system for the first time. The flexibility of the test board with many test points allowed us to have a better understanding of the behaviors of these ASICs. Our study provided key parameters needed for the operation of the electronics and demonstrates the calibration procedure used to find those parameters. The experience we gained will play important roles on the final sTGC FEB design and testing, as well as on the integration and commissioning of FEBs on chambers.

10
**Acknowledgments**

The authors would like to thank the members of the ATLAS NSW collaboration for many useful discussions. The USTC and UM personnel are supported by the National Natural Science Foundation of China (Grant No. 11461141010 and 11875249) and the U.S. Department of Energy (Grant No. 302986 and 326127), respectively.